\documentclass[
 aip,
twocolumn,nofootinbib,
 amsmath,amssymb,
 reprint,%
]{revtex4-1}

\usepackage{graphicx}
\usepackage{dcolumn}
\usepackage{epsfig}
\usepackage{bm}
\usepackage{float}
\usepackage{array}
\usepackage{caption}
\usepackage{subcaption}

\restylefloat{table}

\usepackage{multirow}
\usepackage{color}

\begin{document}

\preprint{AIP/123-QED}

\title{Secondary fast reconnecting instability in the sawtooth crash}  

\author{D. Del Sarto}
\email{daniele.del-sarto@univ-lorraine.fr}
\affiliation{ Institut Jean
Lamour, UMR 7198 CNRS - Universit\'e de Lorraine, F-54506 Vandoeuvre-l\'es-Nancy, France}

\author{M. Ottaviani}
\email{Maurizio.Ottaviani@cea.fr}
\affiliation{CEA, IRFM, F-13108 Saint-Paul-lez-Durance, France}

\date{\today}

\begin{abstract}

In this work we consider magnetic reconnection in thin current sheets with both resistive and electron inertia effects. When the current sheet is produced by a primary instability of the internal kink type, the analysis of secondary instabilities indicates that reconnection proceeds on a time scale much shorter than the primary instability characteristic time. In the case of a sawtooth crash, non-collisional physics becomes important above a value of the Lundquist number which scales like $S \sim (R/d_e)^{12/5}$, in terms of the tokamak major radius $R$ and of the electron skin depth $d_e$. This value is commonly achieved in present day devices. As collisionality is further reduced, the characteristic rate increases, approaching Alfv\'enic values when the primary instability approaches the collisionless regime.   
\end{abstract}

\pacs{Valid PACS appear here}
\keywords{Suggested keywords}
\maketitle


\section{Introduction}

Current sheets are a common occurrence in laboratory and space plasmas. They appear as a local two-dimensional current concentration and they can be either the natural product of convective, often turbulent, plasma motion or the result of a plasma instability entering its non-linear stage. 

In this work, we study the instability of generic current sheets, addressing the question of whether reconnection can be fast enough to account for the observed rates of common events such as 
the sawtooth crash in a tokamak and possibly solar flares.

We consider a plasma characterised by a non-uniform average magnetic field $B_0$, varying on a characteristic scale length $L_0$. The structure of a current sheet is an almost planar current concentration of extension $L \sim L_0$ in two spatial dimensions and thickness $a \ll L$. By definition, the current density in the sheet would differ appreciably from the ambient current density. However one can argue, and it will be shown below, that such current concentration is unstable to fast growing perturbations. Therefore the current density in the sheet $J_{cs}$ cannot exceed too much the ambient current density $J_0$, and therefore one can reasonably assume $J_{cs} \sim J_0 \sim B_0/L_0$.

Moreover, a current sheet would generally evolve in time. However, when studying its stability, it is convenient to assume it to be in a quasi-stationary state, if one is interested in instability growth rates much faster than the underlying evolution rate.
As a consequence, a current sheet can be modelled as a planar system in local force balance. In the following, we further assume that the ambient magnetic field is almost constant and that the current density in the sheet is almost aligned to it. 

The current in the sheet produces a magnetic field $B_{cs} \sim J_{cs} a$, which lays in the plane and it is transverse to the main magnetic field. From the previous estimate of $J_{cs}$ one concludes

\begin{equation}\label{eq:current}
 B_{cs}\sim {B_{0}}\frac{a}{L_0}.
\end{equation}    

This scaling is certainly pertinent when the current sheet is the result of a large $\Delta^\prime$ primary instability as in the case of the sawtooth phenomenon, which is discussed in detail in Sec.~\ref{Sawtooth}. In this instance, scaling (\ref{eq:current}) marks the transition to the nonlinear phase if $a$ is the linear layer width. Therefore we are assuming that the current sheet becomes unstable to secondary instabilities at an early stage of the nonlinear evolution of the primary instability, a result we will find consistent with the estimates derived later in this work. 

Note also that, by assuming this scaling of $B_{cs}$ with the current sheet thickness, we depart fundamentally from previous works~\cite{Jemella,Loureiro1,Loureiro2,Daughton,Uzdensky,Fulvia,Anna_visc,Anna_res,Landi,ideal_de}, where the scaling $B_{cs}\sim B_{0}$ is implicitly assumed. The implications of this different assumption are discussed more in detail in Sec.~\ref{Discussion}. Here we anticipate that
\emph{Pucci and Velli}\cite{Fulvia} pointed out that by increasing the current sheet aspect ratio, tearing mode theory predicts an increasing growth rate, reaching ``\emph{ideal tearing}'', or the the Alfv\'enic limit, at a characteristic aspect ratio that depends on the regime considered\cite{Fulvia,Anna_visc,Anna_res,Landi,ideal_de}. 
The dependence of the growth rate on the aspect ratio is exploited in an essential way also in the present work, as discussed later.

Moreover, the presence of a strong guide field $B_0$ allows one to treat the problem in the framework of reduced MHD.
In this respect, we depart 
from the work of Ref.~\onlinecite{Cassak}, where the guide field is assumed zero. 

In the following, we consider the resistive case, described by the model

\begin{equation}\label{eq:1}
\frac{1}{c_{_A}^2}(\partial_t \nabla_\perp^2 \hat{\phi}
+\hat{\bm u}\cdot{\bm\nabla} \nabla_\perp^2\hat{\phi}) =
\nabla_{||}\hat{J}_{||},
\end{equation}
\begin{equation}\label{eq:2}
\partial_t \hat{\psi}
+\hat{\bm u}\cdot{\bm\nabla} \hat{\psi} 
= -\hat{\eta}\hat{J}_{||},
\end{equation}
 
written in terms of the normalised stream function $\hat{\phi}\equiv c\phi/B_0$, where $\phi$ is the electrostatic potential, and of the normalized magnetic flux function
$\hat{\psi}\equiv \psi/B_0$. Here the current is given by $\hat{J}_{||}\equiv -\nabla^2_\perp \hat{\psi}$, 
the velocity is $\hat{\bm u}= {\bm b_0} \times {\bm\nabla} \hat{\phi}$ and the magnetic field is $\hat{\bm B}={\bm b_0} - {\bm b_0} \times {\bm\nabla} \hat{\psi}$, with $\bm b_0$ the unit vector of the local ambient magnetic field $\bm B_0$. Moreover 
we have introduced the Alfv\'en velocity 
$c_{_A}^2\equiv B_0^2/(4\pi m_i n_0)$ and the magnetic diffusivity 
$\hat{\eta}\equiv c^2\eta/(4\pi)$. The symbol $||$ refers to the parallel component with respect to the total magnetic field and $\perp$ refers to the components perpendicular to the local ambient field. Their exact expression depends on the context and they will be
later specified in the case of slab and cylindrical geometry. Note that in Eqs.(\ref{eq:1}-\ref{eq:2}) \emph{lengths and times still appear in dimensional units}.

In the following  we first derive the scaling of fast reconnection in a general large aspect ratio slab (Sec.\ref{rescaling}). Then, we apply the result to the sawtooth problem, by evaluating whether the sawtooth crash can result from the instability of the current sheet generated by a primary $m=1$ mode in a cylindrical tokamak  (Sec.\ref{Sawtooth}). In particular, we compare our conclusions with the existing numerical results (Sec.\ref{Secondary}), and we analyse the transition to the non-collisional regime. We finally point out the  
difference with other works on fast reconnection, and we outline possible future work (Sec.\ref{Discussion}).


\section{Scaling of the maximum reconnection rate}\label{rescaling}

Labelling with ``$eq$'' the equilibrium quantities and leaving the perturbations unlabelled, we start by linearizing Eqs.(\ref{eq:1}-\ref{eq:2}) around an equilibrium with $\hat{\phi}_{eq}=0$ (no flow),
\begin{equation}\label{eq:lin_1}
\frac{1}{c_{_A}^2} \partial_t \nabla_\perp^2 \hat{\phi} = \nabla_{\parallel eq} \hat{J}_\parallel + {\nabla}_\parallel \hat{J}_{\parallel  eq},
\end{equation}
\begin{equation}\label{eq:lin_2}
\partial_t \hat{\psi} + \nabla_{\parallel eq} \hat{\phi} 
= -\hat{\eta} \hat{J}_{||}.
\end{equation} 
where ${\nabla}_\parallel
=\hat{\bm B}\cdot{\bm\nabla}$.

Adopting a local Cartesian coordinate system, we indicate with $x$ the coordinate perpendicular to the sheet, $y$ the direction of the transverse magnetic field $B_{cs}$ and $z$ the direction of the ambient field $B_0$. Moreover, for
the sake of simplicity
one can assume a symmetric current distribution, such that $B_{cs} = 0$ at $x=0$ and two dimensional perturbations such that $\partial_z=0$.

In the neighbourhood of $x=0$ the parallel gradient operator takes the form
\begin{equation}
\label{eq:grad_par}
\nabla_{|| eq}\sim (B_{cs}^\prime(0) / B_0 ) \, k x \sim (k x/L_0).
\end{equation}   
 
We can now derive the scaling of the growth rate $\gamma$ and of the inner reconnecting layer width $\delta$ in both the tearing (labelled as ``$T$'') and internal kink (labelled as ``$K$'') regimes. 
We recall that these regimes are identified by the conditions $\Delta^\prime \delta < 1$ and $\Delta^\prime \delta > 1$, respectively, where $\Delta^\prime$ is the usual tearing mode stability parameter~\cite{FKR}. 
The scaling can be obtained, up to numerical constants, without carrying out the detailed asymptotic matching calculations, by taking $x \sim \delta$, so that   
$\nabla_{|| eq}\sim (k\delta/L_0)$. Moreover using the fact that $k\ll \delta^{-1}$, $\partial_x \sim \delta^{-1}$ and the estimate $\hat{\phi}''\sim \hat{\phi}/\delta^2$, we obtain the following heuristic balance relations, respectively from  Eq.(\ref{eq:lin_1}) (l.h.s. and first r.h.s. term) and from Eq.(\ref{eq:lin_2}),

\begin{equation}\label{eq:balance_slab}
\frac{1}{c_{_A}^2}\frac{\gamma}{\delta^2}\hat{\phi}\sim 
\frac{k\delta}{L_0} \partial_x^2\hat{\psi},\qquad
\gamma\hat{\psi}\sim \frac{k\delta}{L_0}\hat{\phi}\sim\hat{\eta}\partial_x^2\hat{\psi}.
\end{equation}

These can be now specialised by estimating the perturbed current density respectively as $\partial_x^2\hat{\psi}\sim \hat{\psi}/\delta_{_K}^2$ for the internal-kink (large-$\Delta'$ regime) and as $\partial_x^2\hat{\psi}\sim \hat{\psi}\Delta'/\delta_{_T}$ for the tearing mode (small-$\Delta'$, constant-$\psi$ regime). 

Introducing the macroscopic Alfv\'en time $\tau_0\equiv L_0/c_{_A}$ and the Lundquist number referred to the macroscopic scales $S_0\equiv L_0c_{_A}/\hat{\eta}$, the following scaling is obtained:
  
\begin{equation}\label{eq:slab_tear}
\gamma_{_T}\tau_{0}\sim
S_0^{-\frac{3}{5}} 
\left( kL_0\right)^{\frac{2}{5}}
\left( \Delta'{L_0}\right)^{\frac{4}{5}} ,
\end{equation}
\begin{equation}\label{eq:slab_tear_delta}
  \frac{\delta_{_T}}{L_0}\sim
S_0^{-\frac{2}{5}} 
\left( kL_0\right)^{-\frac{2}{5}}
\left( \Delta'{L_0}\right)^{\frac{1}{5}} ,
\end{equation}  
\begin{equation}\label{eq:slab_kink}
\gamma_{_K}\tau_{0}\sim S_0^{-\frac{1}{3}}
\left( kL_0\right)^{\frac{2}{3}} ,
\end{equation}  
\begin{equation}\label{eq:slab_kink_delta}
  \frac{\delta_{_K}}{L_0}\sim
S_0^{-\frac{1}{3}} 
\left( kL_0\right)^{-\frac{1}{3}} .
\end{equation}

Note that these scalings do not depend explicitly on the current sheet thickness $a$. However, in the tearing regime they do depend implicitly on $a$ via $\Delta^\prime$, which also brings in an additional dependence on the wavenumber.  
If the aspect ratio $L/a$ is sufficiently large so that many wave-numbers are excited, a fastest growing mode exists\cite{Priest,Velli,Batta} 
at the wavenumber $k_{_M}(S_0)$	corresponding to the transition between the tearing and the internal kink regime.
This transition occurs when $\Delta^\prime \delta \simeq 1$.

In order to see this, one has to specify the dependence of $\Delta^\prime$ on $a$ and $k$. Here we adopt the expression obtained for the Harris~\cite{Harris} pinch for small $ka$, that is, well above the tearing instability threshold.

\begin{equation}\label{eq:Delta}
\Delta^\prime \sim \frac{1}{k a^2}.   
\end{equation}

Note that the dependence $\sim 1/k$ of $\Delta^\prime$ for small $k$ is fairly common and that, in any case, what follows can be easily generalised to a different\cite{ideal_de} power-law dependence. 

Using Eq.(\ref{eq:Delta}) in Eqs.(\ref{eq:slab_tear_delta},\ref{eq:slab_kink_delta}) we  obtain 

\begin{equation}\label{eq:slab_k_max}
k_{_M}L_0\sim
S_0^{-\frac{1}{4}}  
\left( \frac{L_0}{a}\right)^{\frac{3}{2}}, 
\end{equation}
\begin{equation}\label{eq:slab_gamma_max}
\gamma_{_M}\tau_0\sim
S_0^{-\frac{1}{2}} 
\left( \frac{L_0}{a}\right),
\end{equation}
\begin{equation}\label{eq:slab_delta_max}
\frac{\delta_{_M}}{L_0}\sim
S_0^{-\frac{1}{4}}
\left( \frac{L_0}{a}\right)^{-\frac{1}{2}}.
\end{equation}

As indicated in the introduction, the above results are of practical interest only if the underlying dynamics occurs at a rate lower than the estimate (\ref{eq:slab_gamma_max}).

When the current sheet is the result of a large-$\Delta^\prime$ primary instability of a quiescent plasma, the time scale of its dynamics can be obtained from 
(\ref{eq:slab_kink})
by taking $k \sim L_0^{-1}$. 
This gives $\gamma_{\rm I} \tau_0 \sim S_0^{-\frac{1}{3}}$. The corresponding layer width is an estimate of the current sheet width, $a \sim \delta_{\rm I} \sim L_0 S_0^{-\frac{1}{3}}$. Using this information one can estimate the maximum growth rate of the secondary instability from (\ref{eq:slab_gamma_max}):

\begin{equation}\label{eq:slab_gamma_max_secondary}
\gamma_{_M}\tau_0\sim
S_0^{-\frac{1}{6}} ,
\end{equation}

which is clearly higher than the primary instability growth rate. One concludes that the current sheet of such primary instability will become quickly unstable, 
as soon as the primary current sheet becomes sufficiently thin. 
An application of this mechanism to the sawtooth crash will be given in the next section.

More generally, current sheets occur as magnetic flux tubes are stretched and twisted by plasma advection. In an ideal plasma, this mechanism produces thinner and thinner current sheets. In a turbulent plasma with comparable kinetic and magnetic energy, the characteristic time is the Alfv\'en time $\tau_0$. 
Therefore a current sheet will break up when its instability rate exceeds $\tau_0^{-1}$. 
From (\ref{eq:slab_gamma_max}) one then deduces that the current sheet thickness must be of the order of:

\begin{equation}\label{eq:SP-thickness}
\frac{a}{L_0} \sim 
S_0^{-\frac{1}{2}}.
\end{equation}

Note that this scaling is the same as the commonly quoted Sweet-Parker thickness. However the analogy appears only superficial. This thickness has been obtained here as the necessary condition for the break-up of a sheet formed by Alfv\'enic motion rather than as the result of the nonlinear reconnection process.


\section{Secondary fast reconnection at the tokamak $n=m=1$ resonant surface}\label{Sawtooth}

We now apply the above results to the $m=n=1$ internal kink (IK) instability\cite{} in a large aspect ratio tokamak.
The IK instability is considered an element of the sawtooth cycle\cite{VonGoeler}, an almost periodic oscillation of plasma temperature characterised by a slow growth and a fast collapse. In the framework of resistive MHD, the IK instability is generally considered too slow to account for the observed fast sawtooth collapse, especially in weakly collisional plasmas\cite{Edwards}, so that non collisional effects are called in to explain observations\cite{Wesson,Porcelli,nonlinear_1}.

Here we take a different approach. We consider a sufficiently resistive plasma, so that the IK can be treated with Eqs.~(\ref{eq:lin_1}-\ref{eq:lin_2}), and we explore whether a secondary instability, possibly in the collisionless regime, can be fast enough to explain observations. 


\subsection{Primary internal-kink resistive  mode}\label{Primary}

For the sake of clarity, in this subsection we review the key elements of the internal kink theory.

We consider a large aspect ratio tokamak in the periodic cylindrical approximation for which the poloidal equilibrium magnetic field $B^0_\theta$ is much smaller than the toroidal one, herewith identified by $B_0$. 
We adopt a set of cylindrical coordinates ($r,\theta,\varphi$) for the radial, poloidal and axial direction, respectively, with periodicity in $\theta$ and $\varphi$.  We also recall the expression of the safety factor:

\begin{equation}\label{eq:q_factor}
q(r)\equiv \frac{r B_0}{RB_\theta^0(r)},
\end{equation}

where $R$ is the major radius, the cylinder length being $2 \pi R$.
Looking for perturbations of the form
 $\hat{\psi}, \hat{J}\sim \cos(n\varphi + m\theta)$ ,
$\hat\phi\sim \sin(n\varphi+m\theta)$, with the replacement $\partial_t\rightarrow \gamma$
and using the corresponding cylindrical expression for the parallel gradient operator 
$\nabla_\parallel = \frac{1}{R}\frac{\partial}{\partial\varphi} + {\hat{\bm \varphi}} \times {\bm\nabla} \hat{\psi} \cdot {\bm\nabla}$ 
we can write the outer equations of the boundary layer problem as
 
\begin{equation}\label{eq:outer_psi}
0=\left(n+\frac{m}{q(r)}\right)
\left[\frac{m^2}{r^2}\hat{\psi} -\frac{1}{r}\frac{\partial}{\partial r}
\left(r \frac{\partial}{\partial r}\hat{\psi}\right)
\right]
\end{equation}
$$+m\hat{\psi} \frac{\partial}{\partial r}
\left[ \frac{1}{r}\frac{\partial}{\partial r}\left(\frac{r^2}{q(r)}\right)\right]
$$
and
\begin{equation}\label{eq:outer_Ohm}
\gamma\hat{\psi}+\frac{1}{R}\left( n+\frac{m}{q(r)}\right)\hat{\phi}=0.
\end{equation}

One can see that, fixing $m=1$ and assuming a rational surface exists in the plasma
at a position $r_1$ such that $q(r_1)=-m/n=1/n$, the exact solution of this equation (with vanishing boundary conditions) for any $n$ and up to an amplitude is
$\hat{\psi}=r\left[(n+1)/q(r) \right]$, $\hat{\phi}=-\gamma r R $,
inside the $r=r_1$ surface, and
$\hat{\psi}=0$, $\hat{\phi}=0$
outside the $r=r_1$ surface.
This implies that $\Delta'=\infty$ for these modes.

The case $n=1$ is of interest for the sawtooth problem in a tokamak. 

In the neighborhood of the point $r_1$ at which $q=-1$ for $m=n=1$,   contained in the inner reconnecting layer of width $\delta$, we can approximate
\begin{equation}\label{eq:grad_para}
\nabla_{|| eq}\simeq \frac{1}{L_s}\frac{r_1-r}{r_1}\sim \frac{\delta}{L_s r_1},
\qquad L_s\equiv \frac{q_1R}{\hat{s}_1},
\end{equation}
where $L_s$ is the shear length in tokamak geometry, $q_1=1$ by definition and $\hat{s}_1=r_1q'(r_1)/q_1$ is the magnetic shear at $r=r_1$.

By carrying out the same balance as in the slab, large-$\Delta^\prime$ case, one obtains 

\begin{equation}\label{eq:kink_primary}
\gamma_{_I}\tau_{0}\sim S_R^{-\frac{1}{3}}
\left(
\frac{r_1}{R\hat{s}_1}
\right)^{-\frac{2}{3}},
\qquad
\frac{\delta_{I}}{R}\sim S_R^{-\frac{1}{3}}
\left(
\frac{r_1}{R\hat{s}_1}
\right)^{-\frac{1}{3}},
\end{equation}

where in this instance the Alfv\'en time and the Lundquist number are defined using $R$ as a normalisation length.
The above results are identical to the slab case by taking $L_0=R/\hat{s_1}$, $k=r_1^{-1}$ and accounting for the change of normalisation length.


\subsection{Secondary instability and the sawtooth crash time scale }\label{Secondary}

At the end of the linear phase, a current sheet develops around the $X$-point, having the form of an helical ribbon of helicity $(1,1)$ and radius $r_1$. As a first approximation, we treat the problem as a planar sheet of width $\delta_{_I}$, given by the inner layer width of the primary mode $m=n=1$. Since at the end of the linear phase the perturbed current density is comparable to the equilibrium one\cite{nonlinear_1,nonlinear_2}, the magnitude of the transverse magnetic field in the current sheet, $B_{cs,\theta}^{_I}$, can be estimated as 

\begin{equation}\label{eq:secondary_B}
\frac{B_{cs,\theta}^{_I}}{B_0}\simeq\frac{x}{R}, 
\quad\mbox{for}\quad x\leq \delta_{I}. 
\end{equation}

As the IK instability enters the nonlinear phase, the extension of the sheet, $L$, would grow from a few times $\delta_{_I}$, to a a fraction of the circumference of radius $r_1$ (e.g., \emph{Waelbroeck} estimates\cite{Waelbroeck} it to be $\sim 2\pi r_1/3$), so we assume $\delta_{_I} < L < r_1$. 
As the extension grows, higher and higher wave-numbers are progressively destabilised, with the smallest unstable wave-vector given by  $k_{min}\equiv 2\pi/L$. 

We now adapt the results of the previous section to the present context. In the following we call $\delta_{_{II}}$ the layer width of the secondary instability and $\gamma_{_{II}}$ the corresponding growth rate. 
For the regime tearing of the secondary mode we obtain

\begin{equation}\label{eq:secondary_tear}
\gamma^{_{(T)}}_{_{II}}\tau_{0}\sim
S_R^{-\frac{3}{5}} 
\left( kR\right)^{\frac{2}{5}}
\left( \Delta'{R}\right)^{\frac{4}{5}},
\end{equation}
\begin{equation}\label{eq:secondary_tear_delta}
  \frac{\delta^{_{(T)}}_{_{II}}}{R}\sim
S_R^{-\frac{2}{5}} 
\left( kR\right)^{-\frac{2}{5}}
\left( \Delta'{R}\right)^{\frac{1}{5}}.
\end{equation}

By estimating $\Delta^\prime$ as in (\ref{eq:Delta}), with $a$ replaced by the IK layer width $\delta_I$, one obtains 

\begin{equation}\label{eq:secondary_tear_2}
\gamma^{_{(T)}}_{_{II}}\tau_{0}\sim
S_R^{-\frac{1}{15}} 
\left( kR\right)^{-\frac{2}{5}}
\left( \frac{r_1}{\hat{s_1} R}\right)^{-\frac{8}{15}},
\end{equation}
\begin{equation}\label{eq:secondary_tear_delta_2}
  \frac{\delta^{_{(T)}}_{_{II}}}{R}\sim
S_R^{-\frac{4}{15}}
\left( kR\right)^{-\frac{3}{5}} 
\left( \frac{r_1}{\hat{s_1} R}\right)^{-\frac{2}{15}}.
\end{equation}
  
The fastest growing mode has a wavenumber

\begin{equation}\label{eq:k_max_II}
k_{_M}^{_{(II)}} R \sim
S_R^{\frac{1}{4}} 
\left( \frac{r_1}{\hat{s_1} R}\right)^{-\frac{1}{2}},
\end{equation}

which belongs to the available range as $L$ becomes comparable to $r_1$.

The corresponding maximum growth rate is

\begin{equation}\label{eq:gamma_max_II}
\gamma_{_{II}}^{_{(M)}} \tau_0 \sim
S_R^{-\frac{1}{6}} 
\left( \frac{r_1}{\hat{s_1} R}\right)^{-\frac{1}{3}}.
\end{equation}

By integrating the reduced-MHD (RMHD) equations in cylindrical geometry, \emph{Yu et al.}\cite{Yu}  have found that the current layer generated by the nonlinear growth of a primary $m/n=1/1$ magnetic island becomes strongly unstable to secondary tearing modes.

According to our calculation~(\ref{eq:gamma_max_II}), in the secondary island regime, a weak positive dependence of the reconnection time on $S_R$, $\sim S_R^{\frac{1}{6}}$, is expected. This appears consistent with the dependence shown in Fig.~5 of \emph{Yu et al.}\cite{Yu} (black bullets).    

In many middle-size tokamaks, the rate given by Eq.~(\ref{eq:gamma_max_II}) is probably fast enough to account for the observed evolution of the sawtooth crash. 
To assess the validity of the resistive model, one has to compare the predicted rate with the electron collision rate $\nu_e$.
Using the fact that $\hat{\eta} \sim \nu_e d_e^2$, where $d_e$ is the electron skin depth, and ignoring geometrical factors, one can consider two cases.

For the validity of Eq.~(\ref{eq:kink_primary}) (primary instability) one requires $\gamma_{_I} < \nu_e$, which can be recast as
\begin{equation}
\label{eq:val_primary}
\tau_0 \nu_e > \frac{d_e}{R}.
\end{equation}

On the other hand, for the validity of Eq.~(\ref{eq:gamma_max_II}) (secondary instability) one requires $\gamma_{_{II}} < \nu_e$, which gives the more restrictive condition
\begin{equation}
\label{eq:val_secondary}
\tau_0 \nu_e > \left (\frac{d_e}{R} \right )^{\frac{2}{5}}.
\end{equation}

In the intermediate regime
\begin{equation}
\label{eq:intermediate}
\frac{d_e}{R} < \tau_0 \nu_e < \left (\frac{d_e}{R} \right )^{\frac{2}{5}},
\end{equation}
the primary instability can be treated with resistive MHD, while a non-collisional model is required to treat the secondary instability correctly.

Finally, when condition~(\ref{eq:val_primary}) is also violated, also the growth rate of the primary instability depends on non-collisional physics~\cite{nonlinear_1,nonlinear_2,Porcelli}. 

\begin{figure*}
\centering{\epsfig{figure=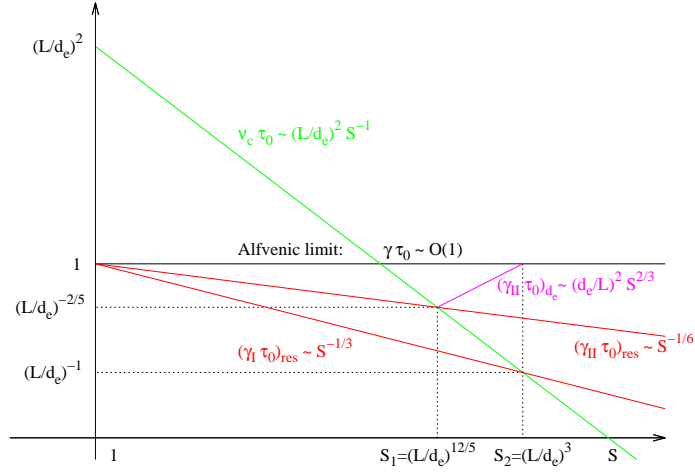,width=9.5cm}}
\caption{Sketch in log-log scale of the maximum growth rate of the secondary instability as a function of the Lundquist number in the collisional (red) and in the non-collisional (mauve) regime. Also shown are the growth rate of the primary instability (also red) and the collision frequency (green).}
\label{regimes}
\end{figure*}

The three collisionality regimes are summarized in Fig.~\ref{regimes}.

The effect of non-collisional physics can be evaluated by considering the model with electron inertia\cite{nonlinear_1} and adapting the estimates of Sec.~\ref{rescaling}. Then, by using again $\Delta^\prime \sim 1/k a^2$, and $R \sim L_0$, Eqs.~(\ref{eq:slab_tear}-\ref{eq:slab_kink_delta}) are replaced by

\begin{equation}\label{eq:slab_tear_ncoll}
\gamma_{_T}\tau_{0}\sim
\left ( \frac{d_e^3 L_0}{a^4} \right ) 
\left( k L_0\right)^{-1},
\end{equation}
\begin{equation}\label{eq:slab_tear_delta_ncoll}
  \frac{\delta_{_T}}{L_0}\sim
\left ( \frac{d_e^2}{a^2} \right ) 
\left( k L_0\right)^{-1}, 
\end{equation}  
\begin{equation}\label{eq:slab_kink_ncoll}
\gamma_{_K}\tau_{0}\sim 
\left ( \frac{d_e}{L_0} \right )
\left( kL_0\right),
\end{equation}  
\begin{equation}\label{eq:slab_kink_delta_ncoll}
  \frac{\delta_{_K}}{L_0}\sim
\frac{d_e}{L_0}. 
\end{equation}

\begin{figure*}
\centering{\epsfig{figure=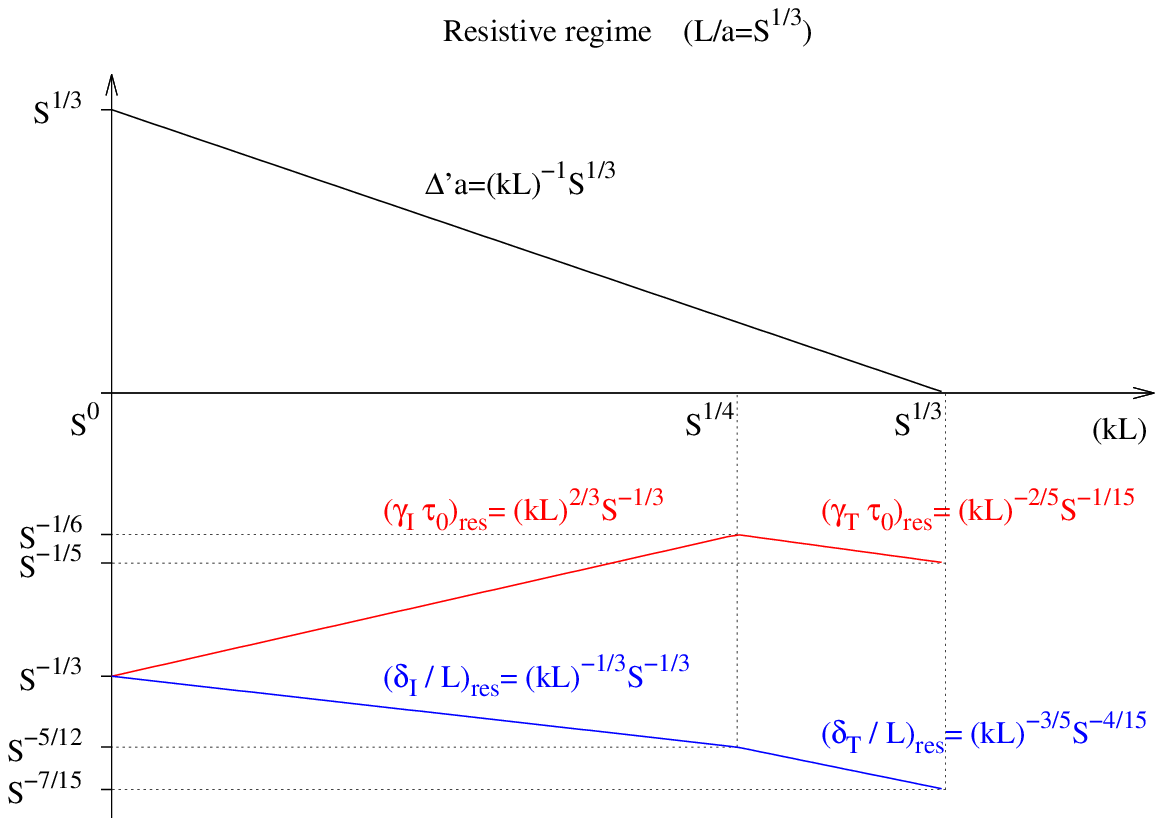,width=8.5cm}
\epsfig{figure=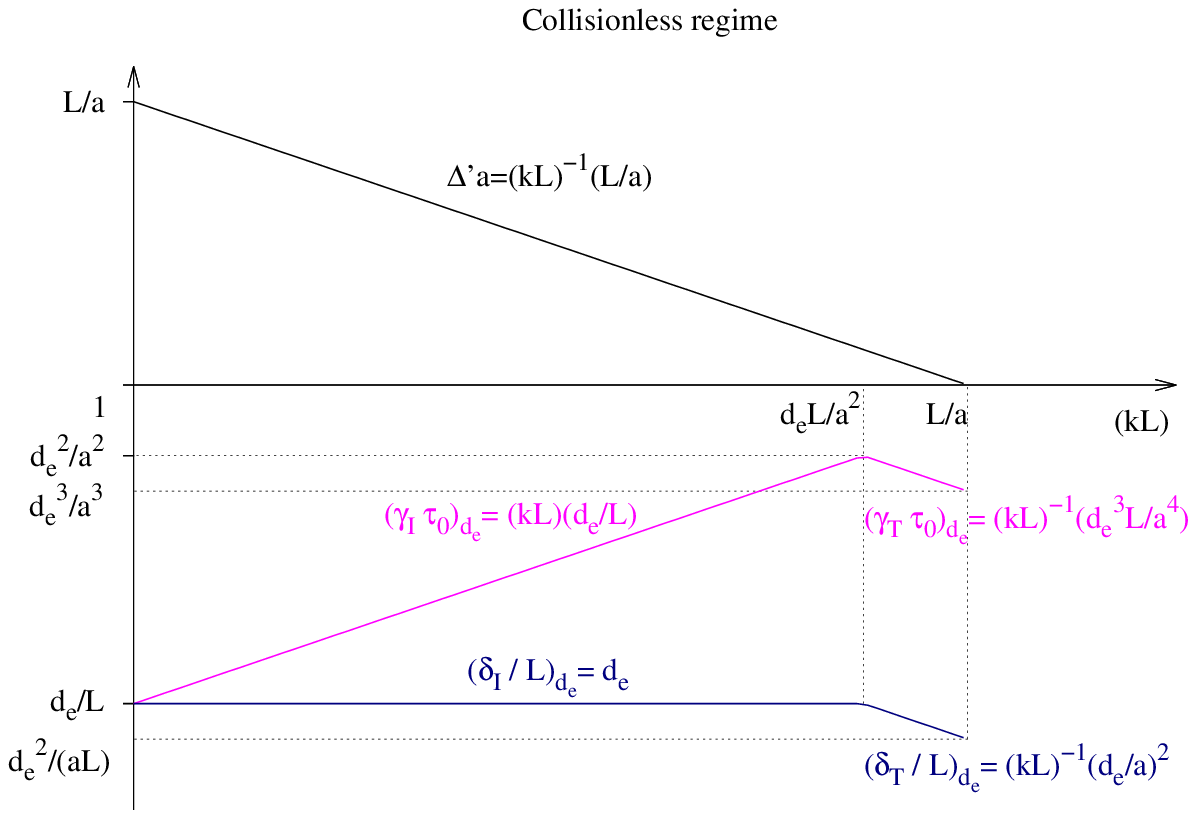,width=8.5cm}}
\caption{Growth rate, layer width, and $\Delta^\prime$ as a function of the wavenumber for the collisional (left) and the non-collisional (right) theory.}
\label{sketches}
\end{figure*}

The growth rate and the layer width as a function of the wavenumber are sketched in log-log plot in Fig.~\ref{sketches} for the collisional (Eqs.~(\ref{eq:slab_tear}-\ref{eq:slab_kink_delta}) with $\Delta^\prime$ given by (\ref{eq:Delta})) and non-collisional regime (Eqs.~(\ref{eq:slab_tear_ncoll}-\ref{eq:slab_kink_delta_ncoll})).

\begin{figure}
\centerline{\epsfig{figure=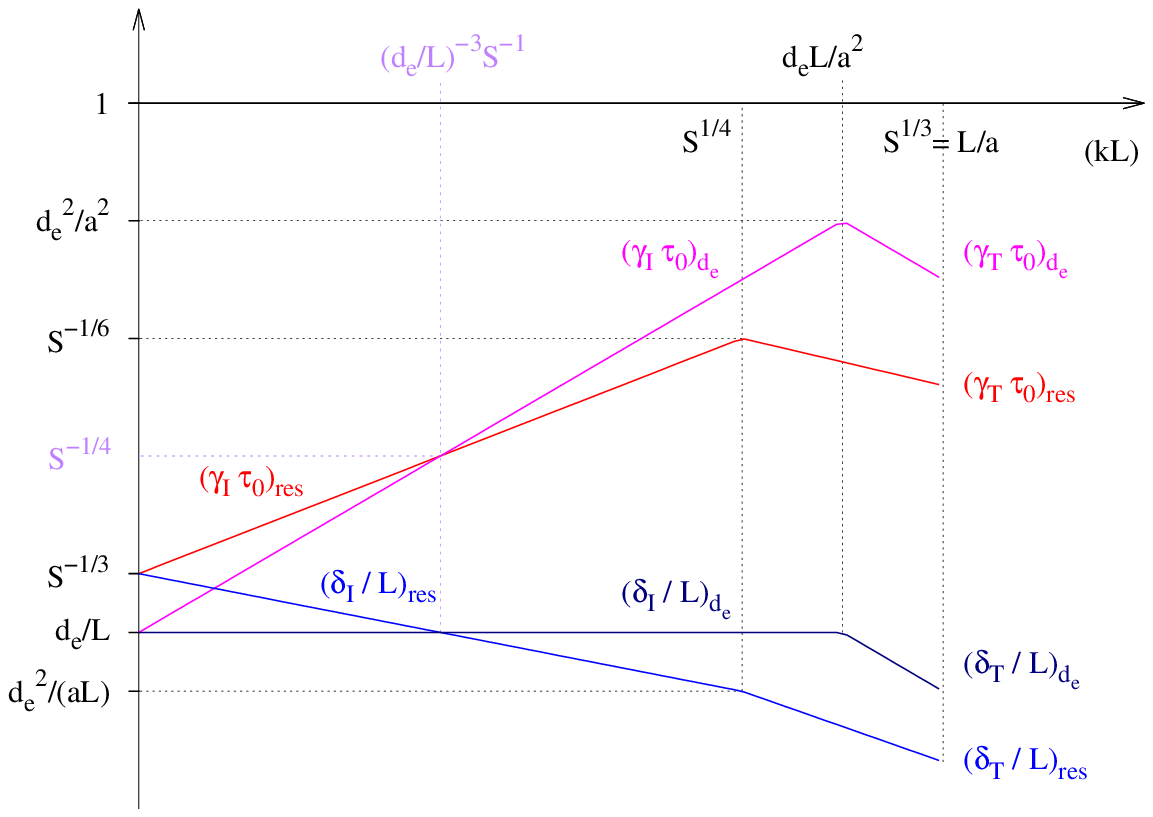,width=9.5cm}}
\caption{ The scalings of the resistive (label ``res'') and of the purely inertia-driven (label ``$d_e$'') regimes at $L=L_0$ and for $(L/d_e)^{12/5}<S<(L/d_e)^3$  are respectively given  by (\ref{eq:slab_kink}) and (\ref{eq:slab_kink_ncoll}) for the primary growth rates ($\gamma_{I}\tau_0$) and by 
(\ref{eq:slab_tear}) and (\ref{eq:slab_tear_ncoll}) for the growth rates of the secondary tearing modes ($\gamma_{T}\tau_0$). Similarly, (\ref{eq:slab_kink_delta}), (\ref{eq:slab_kink_delta_ncoll})  and   
(\ref{eq:slab_tear_delta}), (\ref{eq:slab_tear_delta_ncoll}) give the inner layer widths $\delta_{_I}/L$ and $\delta_{_T}/L$ of the secondary reconnecting modes.}
\label{crossover}
\end{figure}

In the intermediate regime the situation with the two (collisional and non-collisional) estimates is summarized in Fig.~\ref{crossover}.

One can see that, apart from a range at the largest wave-numbers, the growth rate of the secondary instability is determined by non-collisional physics. In particular, the peak growth rate, occurring at the transition between (\ref{eq:slab_tear_ncoll}) and (\ref{eq:slab_kink_ncoll}),
scales like
\begin{equation}\label{eq:slab_gamma_max_secondary_ncoll}
\gamma_{_M}\tau_0\sim
\frac{d_e^2}{a^2}.
\end{equation}  
As one approaches the right boundary of the intermediate regime, the thickness of the primary layer decreases, we recall, as $\sim L_0 S^{-1/3}$. Thus the maximum growth rate of the secondary instability grows again as $\sim S^{2/3}$. 
This behaviour is also shown in Fig.~\ref{regimes}. At the right boundary of this regime, $S \rightarrow (L_0/d_e)^3$ and $a \rightarrow d_e$. Then the current sheet produced by the primary instability approaches the electron skin depth while \emph{the maximum growth rate of the secondary instability approaches Alfv\'enic values}.

The growth rate~(\ref{eq:slab_gamma_max_secondary_ncoll}) is fast enough to account for the observed sawtooth crash time in a machine like JET~\cite{Edwards}. These observations call for a non-collisional theory since the observed rates are faster than the collision frequency. In this respect, referring again to Fig.~\ref{regimes}, we notice that our approach brings the boundary, beyond which the collisionless effects matter, down to $S \sim \left (\frac{R}{d_e} \right )^{\frac{12}{5}}$. This value of the Lundquist number is much more easily achieved in large tokamaks than $S \sim \left (\frac{R}{d_e} \right )^3$ at which the primary instability enters the collisionless regime (upper boundary of the intermediate regime). 
One can also remark that the reconnection rate has a minimum that scales like $\sim (d_e/L_0)^{2/5}$, in Alfv\'en units, in all collisionality regimes.


\section{Discussion and conclusions}\label{Discussion}

The results of \emph{Yu et al.}\cite{Yu} have also been interpreted\cite{Gunter} in terms of the \emph{plasmoid instability}\cite{Loureiro1}. However, it appears that the literature on the plasmoid instability does not take into account the 
rescaling of the current sheet magnetic field with respect to its macroscopic, reference value (Eq.(\ref{eq:current})).
We recall that this is necessary to estimate correctly the size of the current sheet magnetic field at the end of the linear phase of a primary instability such as the $m=n=1$ mode in the sawtooth phenomenon.  
 
By ignoring such rescaling, the secondary instability is faster, such that the maximum growth rate
given in Eq.~(\ref{eq:slab_gamma_max}) would be replaced by\cite{Fulvia}
\begin{equation}\label{eq:slab_gamma_max_norescal}
\gamma_{_M}\tau_0\sim
S_0^{-\frac{1}{2}} 
\left( \frac{L_0}{a}\right)^{\frac{3}{2}}.
\end{equation} 

By assuming that the width of the current sheet resulting from the primary instability scales like $\sim S_R^{-1/3}$, as in this work, together with estimate~(\ref{eq:slab_gamma_max_norescal}), one  
would conclude that the maximum growth rate scales like the \emph{ideal tearing} growth rate $\gamma_{_{II}}\tau_0\sim O(1)$, independent of resistivity.
 
These assumptions are at the root of the work of \emph{Pucci and Velli} as a criterion for \emph{ideal tearing} reconnection rate in a purely resistive model~\cite{Fulvia,Anna_res,Landi}.

If on the other hand one assumed a narrower current layer scaling {\it \`{a}-la} Sweet-Parker, $a \sim L_0 S^{-1/2}$, and at the same time one ignored the rescaling (Eq.(\ref{eq:current})), thereby using again estimate~(\ref{eq:slab_gamma_max_norescal}), one would end up with a faster-than Alfv\'enic estimate of the maximum growth rate $\sim S^{1/4}$. This seems the assumption adopted in the context of the \emph{plasmoid instability} theory~\cite{Loureiro1,Uzdensky,ideal_de}.
 
In the scenario where the current sheet is produced by a primary instability, as in the sawtooth case, we consider 
the results based on the \emph{ideal tearing} and on the \emph{plasmoid instability}
unlikely, since we have shown that the current sheet becomes sufficiently unstable to secondary sub-Alfv\'enic modes at an earlier stage in its development and it would therefore likely break up before becoming sharper, in current concentration, or thinner, in width, than the estimates of Eqs.~(\ref{eq:current})~and~(\ref{eq:slab_kink_delta}), respectively.

In summary, in this work we have shown that 
a current sheet generated by a primary instability, such as the internal kink mode, becomes sufficiently unstable at an early stage in its nonlinear development.

When applied to the context of the sawtooth phenomenon, the analysis of the secondary instability reveals that the sawtooth crash due to reconnection proceeds at a faster rate than previous estimates based on the primary $m=n=1$ internal kink mode~\cite{nonlinear_1}. In the case of a purely resistive model, the reconnection rate depends weakly on the Lundquist number, $\sim S^{-1/6}$, and it appears in agreement with the results of numerical simulations at large Lundquist number~\cite{Yu}. Also, the value of the Lundquist number at which collisionless effects become important is found to be substantially lower than the estimate based on the primary instability only. This broadens the range of tokamaks to which collisionless effects should be taken into account in the analysis of the sawtooth phenomenon. Finally, near-Alfv\'enic reconnection rates can be achieved by secondary instabilities when collisionless effects become important also for the primary instability.    

The scope of this work is to outline a possible scenario for fast reconnection, and in this respect, scaling estimates are useful, and 
sufficient for a first investigation.

Detailed stability analysis of current sheets, taking into account the actual geometry, such as the $m=1$ ribbon, is a possibility to obtain more precise predictions about the maximum growth rate, the associated wavenumber, and the instability threshold. This might lead to an explanation of the Lundquist number threshold observed in numerical simulations~\cite{Yu} and of the number of observed secondary islands. 

The effect of flows also merits an investigation. In this respect, one notes that flows are small near the primary instability X-point, which is a stagnation point, but may be significant far from it, potentially leading to stabilising effects on the tearing mode\cite{Anna_res,Biskamp_2000} and/or to additional instabilities such as Kelvin-Helmholtz's~\cite{DelSarto_KH,DelSarto_KH_2,Loureiro3}.

Finally, the stability analysis carried out in in this work is based on a magnetic configuration with the current aligned to a main magnetic field. Relaxing this hypothesis and allowing for pressure gradients would open further possibilities.\newline
 

\begin{acknowledgments} The authors wish to thank Q. Yu for discussions about his work and A. Tenerani for interesting discussions  and for details  about Ref.\onlinecite{Anna_res}. Discussions with F. Pucci and M. Velli on the ideal tearing are also gratefully acknowledged.
\end{acknowledgments}


\end{document}